**Mastering processing-microstructure complexity through the prediction of thin film structure zone diagrams by generative machine learning models**

**Lars Banko[1], Yury Lysogorskiy[2], Dario Grochla[1], Dennis Naujoks[1], Ralf Drautz[2,3], Alfred Ludwig[1,3]**

[1] Chair for Materials Discovery and Interfaces, Institute for Materials, Ruhr-Universität Bochum, Germany

[2] Interdisciplinary Centre for Advanced Materials Simulation (ICAMS), Ruhr-Universität Bochum, Germany

[3] Materials Research Department, Ruhr-Universität Bochum, Germany

# Abstract

Thin films are ubiquitous in modern technology and highly useful in materials discovery and design. For achieving optimal extrinsic properties their microstructure needs to be controlled in a multi-parameter space, which usually requires a too-high number of experiments to map. We propose to master thin film processing microstructure complexity and to reduce the cost of microstructure design by joining combinatorial experimentation with generative deep learning models to extract synthesis-composition-microstructure relations. A generative machine learning approach comprising a variational autoencoder and a conditional generative adversarial network predicts structure zone diagrams. We demonstrate that generative models provide a so far unseen level of quality of generated structure zone diagrams comprising chemical and processing complexity for the optimization of chemical composition and processing parameters to achieve a desired microstructure.

# Introduction

Thin films are of high importance both in modern technology as they are used as building elements of micro- and nanosystems but also in macroscopic applications where they add functionalities to bulk materials. Furthermore, they play a major role in materials discovery and design [1,2]. Next to composition and phase constitution, the microstructure of thin films is decisive for their properties. The microstructure depends on synthesis conditions and the material itself. Microstructure is important for extrinsic properties, determines functionality and its optimization leads to significant performance enhancement [3–7]. Successful synthesis, e.g. magnetron sputtering, of thin films needs to master many process parameters (e.g. power supply usage (DC, RF, HPPMS), pressure, bias, gas composition, setup and geometry) which determine plasma conditions and affect film growth [8,9]. However, the selection of process parameters, especially for the deposition of new materials, is still mostly based on the scientists' expertise and intuition and these parameters are usually optimized empirically. The film growth and the resulting microstructure at a fixed temperature is primarily determined by the relative flux of all particles in the gas phase, e.g. gas ions, metal ions, neutrals, thermalized atoms, arriving at the substrate [10]. Further, film microstructure is strongly dependent on the energy introduced into the growing surface by energetic ion bombardment [11]. The role of particle-surface interactions in altering film growth kinetics accompanied by thermodynamic mechanisms with respect to microstructure is not yet fully understood.

The need to predict microstructures from process parameters has inspired the development of structure zone diagrams (SZD), first introduced by Movchan and Demchishin for evaporated films [12]. SZD are low-dimensional, abstracted, graphical representations of the occurrence of possible polycrystalline thin-film microstructures (similar structural features) in dependence on processing parameters (e.g. homologous temperature $T_{dep}/T_{melt}$). The simplicity of SZD, which enables estimation of process-dependent microstructures, is also their main drawback, as the actual process parameter space is much larger than what is covered in a classical SZD. Especially with compositionally complex materials, the quality of predictions from simple SZD is limited.

Refined versions of the initial SZD were introduced for magnetron sputtered films: Homologous temperature and sputter pressure [13], homologous temperature and ion bombardment [14], level of contamination [15], reactive gas to metal flux ratio [16], extreme shadowing conditions [17]. Classical SZD for sputtering roughly categorize microstructures into four structure zones (I, T, II and III) [13]. More subzones can be identified based on adatom mobility conditions which influences crystalline texture [18]. Although SZD are useful and popular, they only have a very limited predictive capability since they are based on many generalizations and assumptions, e.g. the pressure is a proxy for the constitution of the incoming particle flux (kinetic energy, ratio of ion-to-growth flux, flux composition). Several revised SZD have in common that they are either strongly abstracted [19] or materials specific [20]. Classical SZD relate processing to microstructure, however only for single elements or binary systems and using system-specific deposition process parameters like gas pressure or substrate bias, which are almost impossible to transfer between deposition systems. In order to identify an ideal microstructure for desired properties, classic SZD are helpful as they give the researchers a hint of likely microstructures, but empirical studies are still required, which require extensive experimental efforts.

To improve the predictive quality of SZDs, multiple input parameters (e.g. incoming particle flux, ion energy, temperature, discharge properties like peak power density and duty cycle, chemical composition, etc.) should be considered conjointly, leading to several challenges, e.g. the visualization of a multidimensional parameter-space. Anders proposed to include plasma parameters and thickness information (deposition, etching) [19]. His SZD keeps three axes, however with two generalized axes (temperature, energy) and the third axis film thickness [19]. However, the generalized axes include unknown factors, i.e. the formula for the calculation of generalized temperature and energy axes. In order to overcome the limitations of SZD, computational methods could be applied. The goal is to achieve a reliable prediction of complex, realistic microstructures based on boundary conditions like composition and relevant process parameters. Microstructures can be predicted by simulations, e.g. kinetic Monte Carlo [21–23] or molecular dynamic simulation [24], which depend on selection of model architectures and boundary conditions and are intractable for high-throughput simulations. The interpretation of the overlap between simulation and experimental results remains to be performed by human assessment. A physical model for an accurate calculation of the microstructure from process parameters needs integrated cross-disciplinary models that cover the plasma discharge at the target, transport of plasma species to the substrate and atomistic processes on the surface and in the volume of the film. Even though progress has been made in different fields (electron [25,26], particle transport [27], plasma-surface-interaction [28], DFT [29]), a unified model is until today out of reach.

If physical models do not exist or are computationally intractable, instead of applying atomistic calculations, machine learning can provide surrogate models bridging the gap between process parameters and resulting microstructure. Machine learning evolved as a new category for microstructure cluster analysis [30,31], microstructure recognition [32–34], defect analysis [35], materials design [36] and materials optimization [37]. Generative models are a class of artificial neural networks that are able to produce new data based on hidden information in training data [38]. The two most popular models are variational autoencoders (VAE) [39] and generative adversarial neural networks (GAN) [40]. VAEs were applied to predict optical transmission spectra from scanned pictures of oxide materials [41], for molecular design [42] and for microstructures in materials design [42–44]. Noraas et al. proposed to use generative deep learning models for material design to identify processing-structure-property relations and predict microstructures [45].

**Our approach**

Many thin films in science and technology have a multinary composition and processing variations lead to an "explosion" of combinations which all would need to be tested to find the best processing condition leading to the optimal microstructure. In order to reduce the cost of microstructure design, we apply machine learning of experimental thin-film SEM-surface images and conditional parameters (chemical composition and process parameters). Two generative models are investigated: a VAE and a conditional GAN (cGAN). The VAE model provides an overview and interpretation of similarities and variations in the dataset by dimensionality reduction and clustering. The generative abilities of the cGAN are applied to conditionally predict microstructures based on conditional parameters. Furthermore, the general ability of deep learning models to generate specialized SZDs based on a limited number of observations is demonstrated. This approach predicts realistic process-microstructure-relations with a generative model being trained on experimental observations only. Our approach handles complexity by (I) performing a limited set of experiments, using "processing libraries" to efficiently generate comprehensive training datasets; (II) training deep learning models to handle SEM microstructure images, (III) visualization of the similarities between different synthesis paths and (IV) predictions of microstructures for new parameters from relations found in the training data. We select a material system from the class of transition metal nitrides, which are applied as hard protective coatings, Cr-Al-O-N [46], for training and evaluation of our models. Cr-Al-O-N and subsystems (e.g. Al-Cr-N, CrN) have been the subject of many studies [47–50]. Our Cr-Al-O-N dataset, efficiently created from materials and processing libraries, in total containing 123 samples, includes variations of six conditional parameters, covering different combinations of compositional (Al-concentration (Al), O-concentration (O) in $Cr_{1-x}$-$Al_x$-$O_y$-N) and process parameters (deposition temperature ($T_d$), average ion energy ($E_I$), degree of ionization ($I_d$) and deposition pressure ($P_d$)). $I_d$ is a design parameter which is related to the ratio of ion flux and the total growth flux of all deposited particles. In order to provide a sufficient quantity of data, 128 patches with size 128 x 128 $px^2$ were extracted randomly from each SEM image (see methods). All depositions were carried out in one sputter system (ATC 2200, AJA International), therefore the geometrical factors that usually change between different deposition equipment is not present. As thin film microstructure is also thickness dependent, all analyzed samples are in a similar thickness range (800 nm - 1300 nm) and exhibit a fully developed microstructure.

To be able to study synthesis-processing-structure relationships, usually a large number of synthesis processes need to be carried out to create a sufficiently large dataset, which is time consuming. To substantially lower the number of necessary synthesis processes, we use combinatorial sputtering of thin-film materials libraries. We introduce the concept of "processing libraries" (PL): These are comparable to materials libraries, but, instead of a composition variation, PL comprise thin films synthesized using a set of different synthesis parameters, at either a constant materials composition, or additionally for different compositions (see methods). The samples in a PL are subject to predetermined variations of the conditional parameters ($E_I$, $I_d$, $T_d$, $P_d$, Al, O). The film growth develops to a microstructure, which is characterized by geometrically different surface features in terms of size, shape and density. For a comprehensive study of possible microstructures, we exploit the process parameter space for synthesis conditions, where either thermodynamic, kinetic or both processes are dominant and repeat these processes for different chemical compositions. Film microstructures are usually assessed by surface and cross-sectional SEM images. Since high quality cross-sectional images are experimentally expensive and their interpretation is complicated, we focus on topographic surface images, as these are more comparable and describable. Surface morphology in terms of grain size and feature shapes can be used to correlate growth conditions and surface diffusion processes with resulting crystallographic orientation [18].

## Results and discussion

In order to inspect the dataset, we train a VAE with a regression model that uses the sampling layer (z) of the VAE as an input to predict the conditions (see methods). The model optimizes simultaneously on microstructure images and conditional parameters and achieves a well-structured and dense representation (latent space embedding). The 64-dimensional latent space is further dimensionally-reduced by kernel principle component analysis (kPCA) with a radial basis function (RBF) kernel [51] in order to provide graphical visualization in 2D. If the microstructure, composition and process parameters correlate, the images should cluster in the VAE latent space.

Figure 1 shows the first two components of the kPCA latent space representation of the validation set. The axes (kPCA 1, kPCA 2) have no actual physical meaning: they are rather a rough expression of how the VAE recognizes images and the conditional parameter space and joins them in a dense layer. Each microstructure image is plotted at its position in the dimensionally reduced latent space embedding of the VAE. The images cluster in regions of similar sizes and shapes. A globular surface morphology is observed at kPCA 1 = -0.1 and kPCA 2 = -0.3. With increasing kPCA 1 and kPCA 2 the feature size decreases. With values of kPCA 1 < 0, mainly facetted grains are observed, while for kPCA 1 > 0 the features become more fine-granular and nanocrystalline.

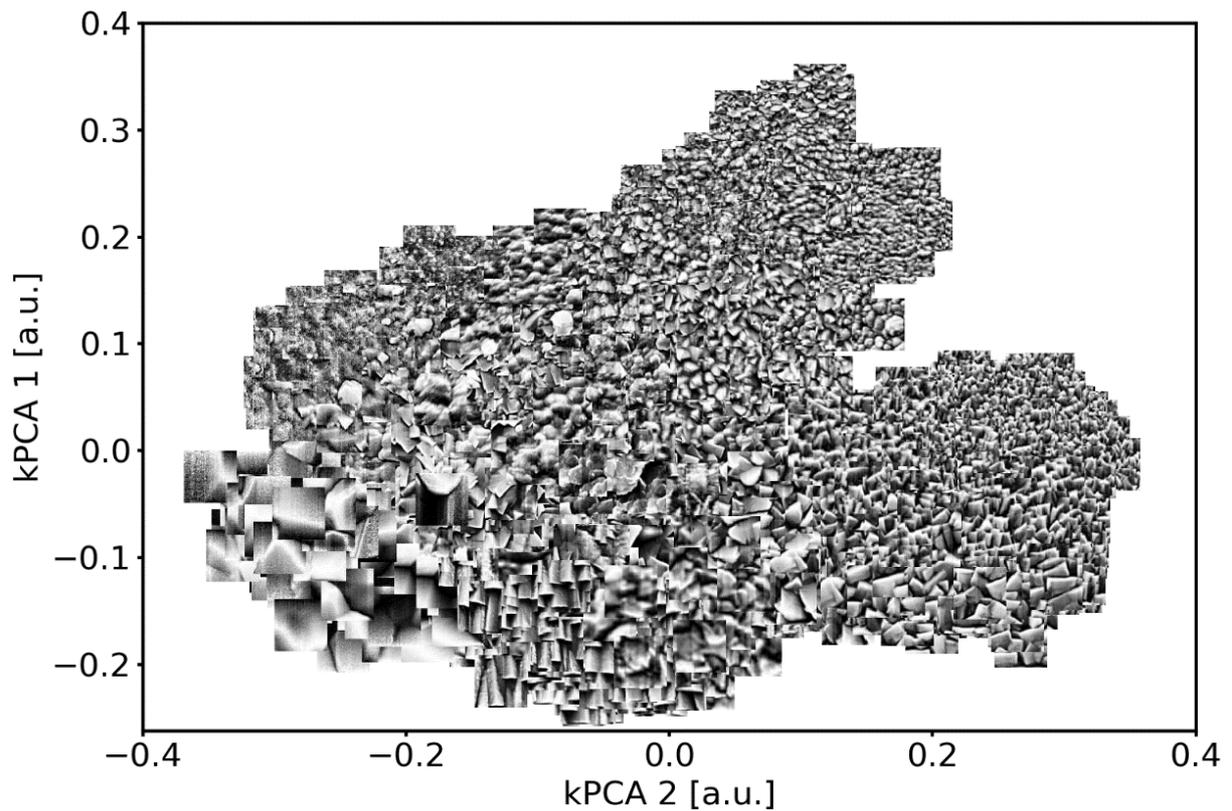

*Figure 1: Latent space representation of all microstructures from the validation set. Patches created from experimental SEM surface images are plotted at their position in the dimensionally-reduced latent space. A continuous variation of microstructures is observed with respect to similarity of feature size and shape (e.g. featureless, nanocrystalline, facetted, smooth facetted, globular).*

**Process-composition-microstructure relations**

This qualitative overview of the microstructures in the dataset is now correlated to chemical composition and process characteristics: Figure 2 shows the microstructure images plotted at their latent space position and the position of each sample in the latent space with their respective color-coded composition or process parameters. This visualizes the interplay between conditional parameters and their significance on microstructural features.

We now address the effect of each deposition parameter in order to provide a discussion baseline for the trends that are created by the prediction of the cGAN model. Samples with different levels of O-contamination are separated in latent space and show a clear trend in feature size (Figure 2e). This can be explained, as O-impurities produce defects in the fcc lattice of Cr-Al-N, inhibiting crystal growth [16,52]. Figure 2c) shows a similar trend for Al [53]. A solid solution for $Cr_{1-x}$-$Al_x$-N with up to 70 at. % Al is known [49], whereas between 50 and 70 at.% Al, hcp AlN precipitates [54]. The maximum solubility of Al in fcc CrN depends on process parameters [55]. An increase in $T_d$ (Figure 2b) leads to an increase in feature size. The feature shapes change from fine granular to facetted grains and at high $T_d$ to globular grains due to higher diffusion rates [20]. An increase in $E_l$ (Figure 2d) leads to a smoother surface as kinetic bombardment flattens facets and in extreme cases a featureless surface is observed. Additionally, surface diffusion is kinetically enhanced by ion bombardment [56]. An increased $I_d$ (Figure 2f) leads to a more directed particle flux resulting in oriented facets when the particle flux is inclined to the substrate normal [8]. In our case, the two confocal cathodes are inclined by 27° to the substrate normal to achieve a composition gradient. An increasing $P_d$ (not shown) leads to an increase in gas atoms or molecules per volume and thereby to a decrease in mean free path [57]. Particles experience more collisions during their path from the target surface to the substrate and thereby lose energy. Additionally, $I_d$ and the ratio of gas ions to target ions increases, which influences surface kinetics. This illustrates the complex interplay between process parameters, composition and resulting microstructure. Also, it shows the usefulness of dimensionality reduction to gain an overview of complex datasets. The identified trends correlate well with results from literature.

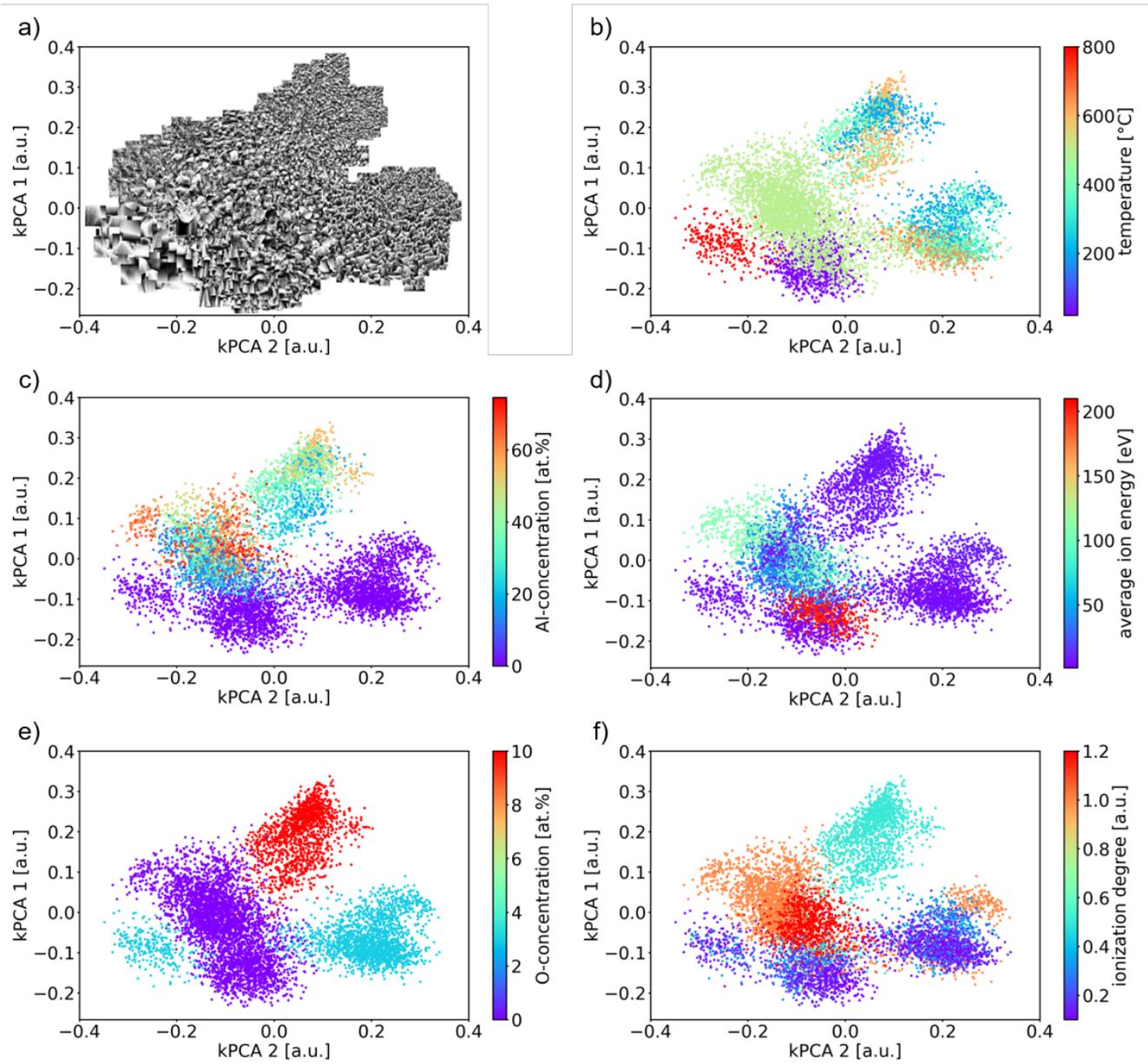

Figure 2: Correlation of the microstructures in the dataset with chemical composition and cconditional parameters. VAE latent representation of microstructures (a) and conditional parameters (b) $T_d$, c) $A_l$, d) $E_I$, e) O, f) $I_d$) from the validation set.

**Prediction of microstructures from conditional parameters**

The decoder part of the VAE could be applied to generate images from the latent representation, but the quality is unsatisfactory due to known limitations of VAEs [58]. In contrast, GAN models are known to be able to produce photorealistic images [59]. To predict microstructures from the six conditional parameters, we train a cGAN model [60]. In order to categorize the level of prediction, we need to define what the model can learn from the experimental dataset. A reconstruction of a microstructure from the training set provides the baseline. Figure 3 compares experimental images to their predicted counterparts. The cGAN generates these microstructure images using two inputs only: conditional parameters and a latent sub-space with random noise. It should be noted that the cGAN is not trained to generate an exact copy of the original image. The generated images generally show a good reproduction of the experimental images in terms of feature size and shape. Even contrast variations on facets are reproduced. Figure 3 g) shows an exception, where locally, smaller features are generated on top of otherwise large smooth grain surfaces. This relates to the problem that the image patches only show small fractions of these large grains and the microstructure of 800°C deposited samples strongly differs from all other images. In rows b), c) and f) the generated images are nearly indistinguishable from their experimental counterparts. The facet shapes in row a) are not as sharp as in the experimental images and the facets in row d) show more curvature compared to the original images. However, the reproduced features can still be identified as facetted and the feature sizes match well. The generated images in row e) appear blurred and the feature size is smaller than the experimental image. A low contrast in the experimental images of these smooth dense microstructures might affect the training of the model.

| | conditional parameters | | | | | original images | conditionally generated images |
|---|---|---|---|---|---|---|---|
| | $T_d$ [°C] | Al [at.%] | O [at.%] | $I_d$ [a.u.] | $E_I$ [eV] | | |
| a) | 25 | 0 | 0 | 0.2 | 2 | | |
| b) | 200 | 17 | 10 | 0.5 | 8.5 | | |
| c) | 300 | 0 | 3 | 0.1 | 3 | | |
| d) | 500 | 20 | 0 | 1 | 42 | | |
| e) | 500 | 0 | 0 | 1 | 200 | | |
| f) | 600 | 27 | 10 | 0.5 | 7 | | |
| g) | 800 | 0 | 3 | 0.2 | 5 | | |

240 nm

*Figure 3: Comparison of randomly-selected experimental SEM surface microstructure images and cGAN-generated images for the same conditional parameters.*

Machine learning models can only learn from information provided to them. Interpolations are reasonable while extrapolations are more challenging. For example, a microstructure prediction for a sample with high Al at $T_d$ = 1000°C would fail, as a phase decomposition is expected which leads to a (for the model) unpredictable microstructure. The training set libraries were synthesized at selected basic process conditions (e.g. constant $T_d$, $E_I$, Al or O) and contain a variation of one or two additional parameters. Therefore, the complete dataset has only limited intersections and extensions into other conditions. E.g., a variation of $E_I$ between 40 eV and 200 eV was only carried out for samples deposited at 500°C. In order to predict a sample deposited at $T_d$ = 100°C and $E_I$ = 200 eV, a transfer of the $E_I$ trend on the $T_d$ trend is required. To validate the predictive capabilities of the model, new microstructure images are generated for extensions of a chosen base condition. A CrN sample ($T_d$ = 20°C, $E_I$ = 1 eV, $I_d$ = 0.1, $P_d$ = 0.5, Al = 0 at.%, O = 0 at.%) provides the base condition (Figure 4, orange frame). The sample exhibits a triangular facetted morphology. Figure 4 visualizes how the microstructure of the initial sample changes when only a single condition is changed at a time and the other parameters stay constant. Experimental structures synthesized with the same or similar process parameters (closest experimental condition) are compared. The trends from Figure 2 for the different conditional parameters are reproduced by the model. In the example Al and O lead to a refinement of the microstructure, $E_I$ leads to smoother facets, $I_d$ increases the orientation of the grains and $T_d$ increases the grain size.

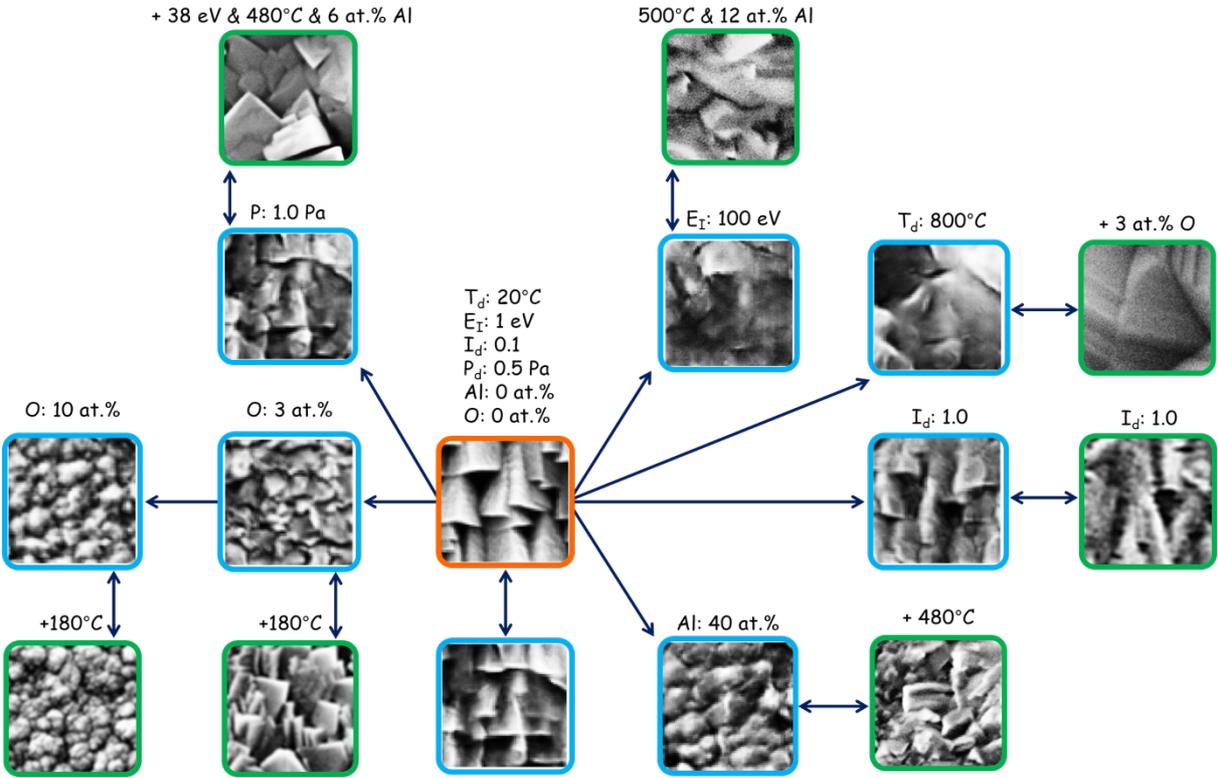

*Figure 4: Predicted microstructures from cGAN (blue frame) and comparison with experimental results (green frames). Predictions are based on the microstructure of Cr-N deposited at room temperature (orange frame) with corresponding conditional parameters. The changed parameter for each predicted or experimental image is indicated while all other parameters stay constant. If an experimental counterpart with identical conditional parameters is missing, the closest related experimental image is selected and the additional change in conditional parameters is marked above the image (e.g. +480°C).*

To validate the cGAN prediction quality, microstructures from experimental test samples which were not included in the training set are predicted. $Cr_{1-x}Al_xN$ samples grown at 500°C from the training set were deposited at < 10 eV (0 V substrate bias) and >100 eV (-100 V substrate bias) for different Al-concentrations. As an example, the cGAN predicts images for different Al at 40 eV (Figure 5). This requires an interpolation of $E_I$. At 0 V bias, the faceted microstructure changes to a fine-grained microstructure with increasing Al. The same trend is observed at -100 V bias but the facets of Cr-rich samples are smoother and denser. With increasing Al, the microstructure becomes featureless. The prediction matches both trends. In direct comparison to the experimental counterparts, the facets of Cr-rich samples are less pronounced. Al-rich samples are almost indistinguishable from the test set images. These results show that the cGAN produces excellent results for interpolations within the data set.

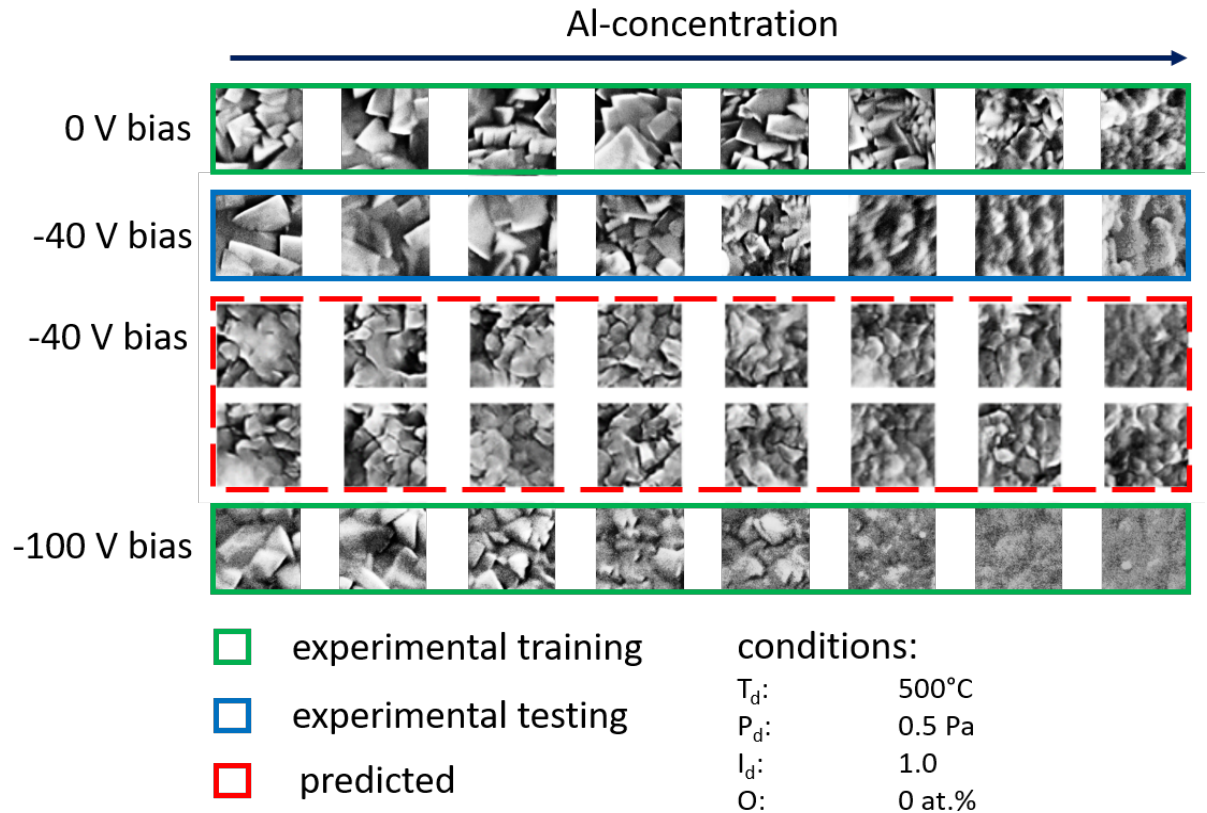

*Figure 5: Synopsis of experimental and predicted images. Green boxes contain experimental images from the training set. The blue box contains experimental images from the test set. The red box contains images predicted by the cGAN for the same conditional parameters of the test set. The data shows the effect of bias voltage ($E_I$) and Al on the surface microstructure.*

Finally, a SZD is generated by the cGAN. The advantage of this generative SZD (gSZD) is that it can be produced as required. In a 2D representation, two parameters can be varied while the remaining four parameters are selected constant. Figure 6 shows a gSZD for a variation of Al and $T_d$ at constant values for the remaining parameters. Al and $T_d$ are varied randomly between 0 to 70 at.% and 20 - 800°C, respectively. The predicted image patches are plotted at positions according to their input conditions. Hence, patches overlay and appear as a continuous diagram. A clear variation of the microstructure in dependence of $T_d$ and Al is observed. The structure changes with increasing Al from facetted to a smoother, fine-grained structure. At $T_d > 350°C$, Cr-rich samples exhibit a denser microstructure with smoother grains. Regions are highlighted in the diagram where structure changes were identified. The remaining parameters (O, $I_d$, $E_I$) vary in the experimental data, while they are kept constant in the gSZD. A variation of Al and $T_d$ was experimentally realized at 10 at.% O while samples with a variation of $E_I$ were deposited at 500°C and contain 0 at.% O. Thus, the model combines (I) the structural refinement with increasing Al and (II) the trend that this refinement is inhibited with increasing $T_d$. In other words, a higher $T_d$ is necessary at high Al to obtain a similar feature size and shape as compared to Cr-rich compositions without O and Al. For the gSZD these can be interpreted in the following way. In general, an increase in Al leads to a refinement of the microstructure due to changes in adatom surface mobility conditions [61]. This trend is most significant at low temperatures, since surface kinetics surpass thermodynamic processes. With increasing temperature, the faceted structure extends to higher Al. An increase in feature size with $T_d$ is observed. The observed trends change to a finer structure when O is increased stepwise and facets are smoothed out by a stepwise increase of $E_I$ (gSZD not shown).

**Definition of conditions for thin films with optimized microstructures**

Finally, by combination of domain knowledge and the new gSZD, we are able to design a composition-process-window to create films for a desired application. For an example application of hard protective coatings for polymer injection molding or extrusion tools [62], the tribological performance needs to be optimized, requiring films with a dense, smooth microstructure. Physical boundaries are provided by the maximum values of Al and $T_d$. $T_d$ is limited by the temper diagram of cold work steel AISI 420 (X42Cr13, 1.2083). To avoid tempering of the substrate, the maximum $T_d$ should be lower than 450°C. Al is limited by the formation of hcp AlN above 50 at.% Al, which would lead to a reduction in hardness [63]. To achieve a smooth and dense film, Al should be as high as possible, according to the gSZD. Additionally, $T_d$ should be as high as possible in order to reduce grain boundary porosity. With an included uncertainty, the new composition-process-window ("window of opportunity") is provided by the green region in **Fehler! Verweisquelle konnte nicht gefunden werden.**.

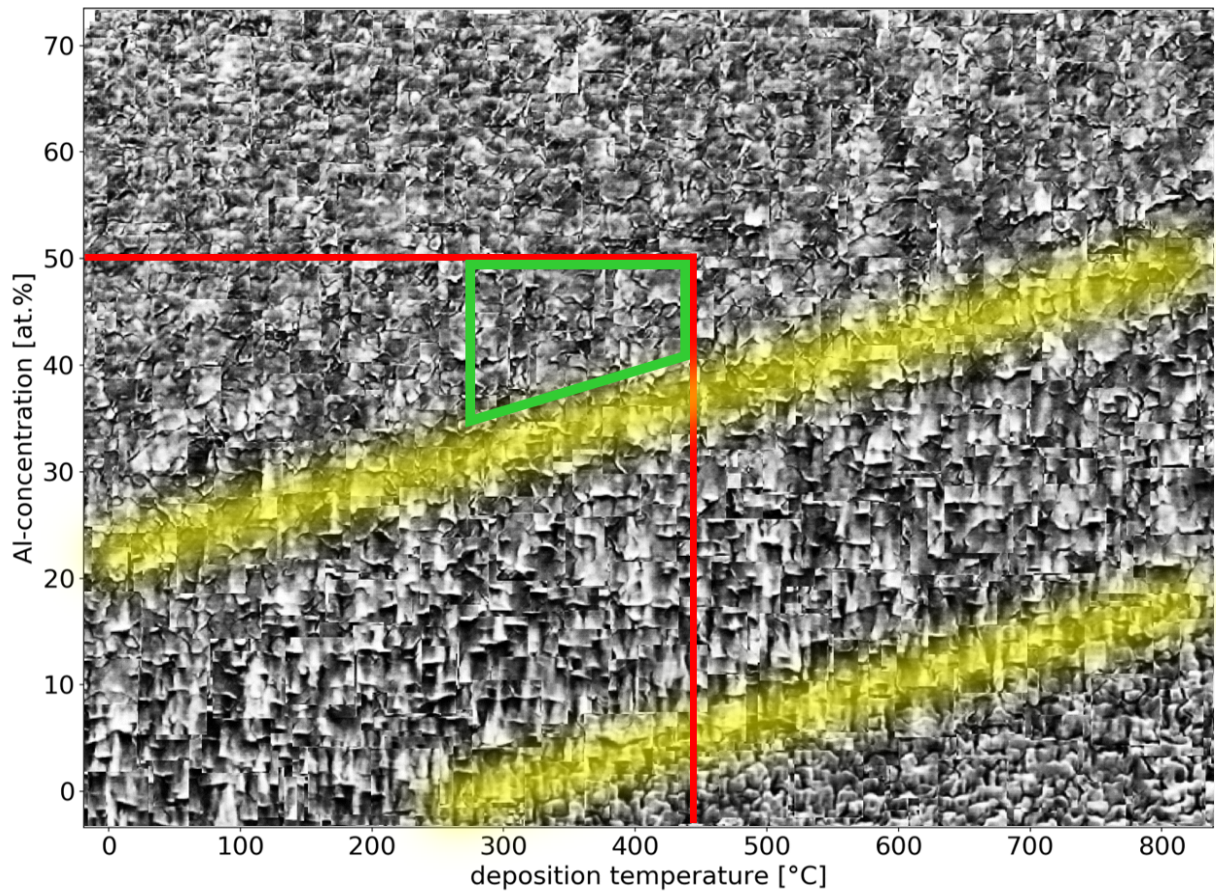

*Figure 6: gSZD generated by the cGAN model for a variation of Al und $T_d$. The remaining process parameters were chosen to be: $I_d$ = 0.1, $E_I$ = 10 eV, O = 0 at%, $P_d$ = 0.5 Pa. The yellow regions highlight the separation of zones with characteristic microstructural features. Red lines indicate the technological boundaries of the use case. The area outlined in green shows a window of opportunity.*

# Conclusion

We applied combinatorial synthesis methods to create materials and process libraries of the Cr-Al-O-N system in order to observe the influence of composition and process parameters on the resulting microstructural properties. Our training set of samples from the Cr-Al-O-N system covers variations in the directions of previous SZD ($T_d$, $P_d$, $I_d$, $E_I$, O) and an additional compositional variation of Al. A generative neural network was trained on SEM surface images to predict microstructures based on the input of composition and process parameters. The model reproduces the observed trends in the dataset. Furthermore, we were able to validate the predictive capabilities on test data, which requires an interpolation of conditional parameters. A transfer of trends from sampled regions to un-sampled regions was demonstrated in a new generative SZD. The gSZD shows the expected microstructure of thin films for a variation of Al concentration and deposition temperature, which will be useful for the optimization of TM-Al-N (TM = transition metal) thin films. A so far unseen level of predictive quality in the scope of SZD is observed which will lead to an acceleration in the development and optimization of thin films with a desired microstructure. Until now, the assessment of new, predicted microstructures needs to be performed by the scientist. A system of metrics for the evaluation of microstructures would be helpful for future works in this field. Nevertheless, the results give reason to believe that future generative models will be applied in the microstructure optimization process, which is a key challenge in functional thin film materials.

# Methods

**Sample synthesis**

Sample synthesis is performed in a multi-cathode magnetron sputter chamber (ATC 2200, AJA International). All samples are deposited reactively with an Ar/N$_2$ flux ratio of 1 and a total gas flux of 80 sccm. The deposition pressure is controlled automatically by adjusting the pumping speed. Two confocal aligned cathodes (Al, Cr) facing the substrate lead to a continuous composition gradient of the two base materials, which results in a materials library. The substrate is heated with a resistive heater. In order to create a PL, an in-house made step heater is used to heat 5 substrates simultaneously at 5 different temperatures in the range from 200°C to 800°C, thereby covering a large temperature range of typical SZDs within a single PL [20]. PLs with a continuous variation of plasma parameters, e.g. $E_I$ and $I_d$ are synthesized by sputtering from two confocally aligned magnetrons which are operated by different power supplies. One cathode is powered by DC, the other one by HiPIMS (high power impulse magnetron sputtering). The substrate is placed centered below the two cathodes. A similar concept was chosen by Greczynski et al. [64]. The pulsed HiPIMS discharge produces a one magnitude larger number of ionized species and higher ion energies compared to a DC discharge [65]. An additional substrate bias is applied in some cases to further accelerate ions and increase $E_I$. By placing the substrate in the center below the two inclined cathodes, the travel distance of the ionized species of the HiPIMS discharge increases towards the substrate positions next to the DC cathode. The ions thermalize due to collisions with other plasma species and loose energy. This effect is amplified by the angular distribution of the sputtered species [66]. Consequently, the ratio of ions per deposited atom as well as the average ion energy are different along the 100 mm diameter substrate. In order to achieve a homogenous film thickness, the DC power is reduced to match the typically lower deposition rate of the HiPIMS powered cathode. A variation in the degree of ionization is achieved by a variation of the sputter frequency at constant average power in HiPIMS processes. An increase in frequency leads to a decrease in target peak power density which leads to a decrease in $I_d$ and a small decrease (up to 3 eV) in $E_I$. The O-concentration in several of the discussed samples are contaminations from residual gas outgassing from the deposition equipment, which is especially present at elevated temperatures (> 600°C).

**Thin film characterization**

The chemical composition (Al/Cr) is determined by EDX (Inca X-act, Oxford Instruments). The O-concentration is determined by XPS (Kratos Axis Nova) for a subset of the samples. All films are stoichiometric in terms by the definition of (Al+Cr)/(O+N) = 1. The stoichiometry is validated for additional samples that are deposited under similar process conditions (not shown) by RBS measurements, within a 5 at.% error. SEM images are taken in a Jeol 7200F using the secondary electron detector at 50,000x magnification at an image size of 1280 x 960 pixels. The SEM images are histogram-equalized using contrast limited adaptive histogram equalization (CLAHE) [68].

**Plasma properties**

$E_I$ was calculated from retarding field energy analyzer measurements of a previous study[67] that were carried out at five measurement positions along the 100 mm substrate area in three reactive co-deposition processes of Al and Cr at 100, 200 and 400 Hz sputter frequency at 0.5 Pa. If a substrate bias was applied, an additional ion energy was added to the total ion energy (e.g. $E_I$ + 40 eV bias). To estimate $I_d$, the ratio of total ion flux and growth flux was calculated. Unknown values for conditions that were not measured are estimated by extrapolation. The ion to growth flux ratios are normalized over the data set. These values provide only a rough estimation that covers the known trends from literature and our own investigations. It should be noted that we consider $I_d$ a physics-informed descriptor, rather than a physical property.

**Data handling**

Our dataset contains 123 individual samples. The 1280 x 960 px$^2$ images locally contain characteristic microstructure features that are distributed repeatedly over the image. Patches are extracted at random points of each image. Each of the extracted patches cover a large enough range to represent the characteristic microstructure of the synthesis condition. We choose a patch size of 128 x128 px$^2$ and scale them by a factor of 2 into 64 x 64 px$^2$ to speed up computations. The images patches have a pixel density of 27 px/nm. A total of 128 patches are cropped per each image which results in an average pixel shift of 10 and 7.5 px per patch (1280/128, 960/128). The training data therefore contains more than 10.000 different image patches depending on the train-test split. For the VAE, the complete data set is split randomly at a ratio of 70:30 (train:validation). In case of the cGAN, a test set (13 out of 123 original SEM images) for the conditions (described in Figure 5) is removed from the data set.

**Machine learning models**

**VAE**

The VAE model consist of 3 models, an encoder, a decoder and a regression model. Encoder and decoder represent the variational autoencoder (VAE) part of the model. The image patches of size (64x64x1) provide the input and the output of the VAE. The encoder consists of 5 convolution building blocks which comprise a 2D convolutional layer that is followed by batch normalization, a Leaky ReLU activation function and a dropout layer. The filter sizes are 32, 64, 128, 128, 128. The kernel size is 4 x 4. The output of the last convolutional layer is flattened and connected to two dense layers (μ and σ) with 64 dimensions. These are passed to a sampling layer (z) which samples the latent space according to the formula: $z = \mu + \alpha \varepsilon e^{\sigma/2}$. ε is a random normal tensor with zero mean and unit variance and has the same shape as μ. α is a constant which is set to 1 during training and otherwise to 0. The decoder reflects the structure of the encoder. The output of the sampling layer is passed into a dense layer with 512 neurons which is reshaped to match the shape of the last convolutional layer. The layer is passed to 5 building blocks which comprise a 2D convolutional layer followed by batch normalization, Leaky ReLU activation, dropout and an upsampling layer. The filter sizes of the convolutional layers are 128, 128, 128, 64, 32. An additional convolutional layer with filter size 1 provides the output of the decoder. A regression model takes the output of the sampling layer z as an input and outputs the conditional parameters. The regression model has 4 dense layers with dimensions 20, 20, 20, 6 and ReLU activation, an input layer with 64 dimensions and an output layer with 6 dimensions and linear activation function. The VAE and the regression model are simultaneously trained using the Adadelta optimizer. The VAE loss is provided by the sum of the Kullback–Leibler divergence and the image reconstruction binary cross entropy. The loss of the regression model is calculated by the mean squared error. The losses of VAE and regression model are weighted 1:10000 in order to provide a well-structured latent space.

**cGAN**

The generative adversarial network consists of two parts: a generator and a discriminator. The generator network has two inputs, a 16-dimensional latent space (intrinsic parameters) and six conditional physical parameters (extrinsic). The latent space input layer is followed by a dense layer with 32768 neurons and Leaky ReLU activation function and then reshaped into a 16x16 layer with 128 channels. The conditional input layer is followed by 256 dense layers with linear activation function and reshaped into a 16 x 16 matrix with one channel. Two reshaped 16x16 matrices are combined together and followed by two convolutional-transpose layers with Leaky ReLU activation functions, with an upscaling factor of 2 and 128 filters for each layer. The last layer is convolutional with hyperbolic tangent activation and 64 x 64 x 1 shaped of output. The discriminator network also has two inputs, the six conditional physical parameters and a 64 x 64 x 1 input image. As in the generator network the conditional input layer is converted into a 64 x 64 x 1 matrix with one dense layer and concatenated with the input image. This is followed by two convolutional layers with 128 channels and a downscaling factor of 2, which results in a 16 x 16 x 128 matrix. A flattening layer is followed by a dropout layer with a dropout factor = 0.4 and a dense output layer with sigmoid activation function. The same conditional extrinsic physical parameters were fed into both the generator and the discriminator. The discriminator model has a binary cross-entropy loss function and an Adam optimizer with a learning rate equal to 0.0002, and beta_1 equal to 0.5. The loss function for the generator is approximated by the negative discriminator, in a spirit of adversarial network training. The training procedure consists of consecutive training of the discriminator on small batches of real and fake images with corresponding conditional physical parameters and generator training on randomly generated points from latent space and realistic extrinsic parameters.

## Acknowledgements

This study was funded by the German Research Foundation (DFG) as part of the Collaborative Research Centre TRR87/3 "Pulsed high power plasmas for the synthesis of nanostructured functional layers" (SFB-TR 87), project C2. The authors thank Alan Savan for cross-reading the manuscript.

## Author contributions

L.B. developed the concept, deposited the samples and adapted the machine learning models, L.B. and A.L. wrote the main parts of the manuscript, Y.L. contributed the code of the machine learning models and coded the cGAN model and helped to develop the machine learning methodology, D.G. and D.N. were involved in preparation and analysis of materials and processing libraries, R.D. supervised the development of the machine learning methods. All authors contributed to the writing of the manuscript.

## Competing Interests

The authors declare no competing interests.

# References

1. Alberi, K. *et al.* The 2019 materials by design roadmap. *Journal of Physics D: Applied Physics* **52,** 13001 (2018).

2. Ludwig, A. Discovery of new materials using combinatorial synthesis and high-throughput characterization of thin-film materials libraries combined with computational methods. *npj Computational Materials* **5,** 70 (2019).

3. Greczynski, G., Jensen, J., Böhlmark, J. & Hultman, L. Microstructure control of CrNx films during high power impulse magnetron sputtering. *Surface and Coatings Technology* **205,** 118–130; 10.1016/j.surfcoat.2010.06.016 (2010).

4. Pan, T. S. *et al.* Enhanced thermal conductivity of polycrystalline aluminum nitride thin films by optimizing the interface structure. *Journal of Applied Physics* **112,** 44905; 10.1063/1.4748048 (2012).

5. Wang, X. C., Mi, W. B., Chen, G. F., Chen, X. M. & Yang, B. H. Surface morphology, structure, magnetic and electrical transport properties of reactive sputtered polycrystalline Ti1–xFexN films. *Applied Surface Science* **258,** 4764–4769; 10.1016/j.apsusc.2012.01.088 (2012).

6. Zalnezhad, E., Sarhan, A. A. D. & Hamdi, M. Optimizing the PVD TiN thin film coating's parameters on aerospace AL7075-T6 alloy for higher coating hardness and adhesion with better tribological properties of the coating surface. *Int J Adv Manuf Technol* **64,** 281–290; 10.1007/s00170-012-4022-6 (2013).

7. Zgrabik, C. M. & Hu, E. L. Optimization of sputtered titanium nitride as a tunable metal for plasmonic applications. *Opt. Mater. Express* **5,** 2786; 10.1364/OME.5.002786 (2015).

8. Depla, D. & Mahieu, S. *Reactive sputter deposition* (Springer, 2008).

9. Kouznetsov, V., Macak, K., Schneider, J. M., Helmersson, U. & Petrov, I. A novel pulsed magnetron sputter technique utilizing very high target power densities. *Surface and Coatings Technology* **122,** 290–293 (1999).

10. Kay, E., Parmigiani, F. & Parrish, W. Microstructure of sputtered metal films grown in high-and low-pressure discharges. *Journal of Vacuum Science & Technology A: Vacuum, Surfaces, and Films* **6,** 3074–3081 (1988).

11. Harper, J. M. E., Cuomo, J. J., Gambino, R. J. & Kaufman, H. R. Modification of thin film properties by ion bombardment during deposition. *Nuclear Instruments and Methods in Physics Research Section B: Beam Interactions with Materials and Atoms* **7,** 886–892 (1985).

12. Movchan, B. A. & Demchishin, A. V. STRUCTURE AND PROPERTIES OF THICK CONDENSATES OF NICKEL, TITANIUM, TUNGSTEN, ALUMINUM OXIDES, AND ZIRCONIUM DIOXIDE IN VACUUM. *Fiz. Metal. Metalloved. 28: 653-60 (Oct 1969).* (1969).

13. Thornton, J. A. The microstructure of sputter-deposited coatings. *Journal of Vacuum Science & Technology A: Vacuum, Surfaces, and Films* **4,** 3059–3065; 10.1116/1.573628 (1986).

14. Messier, R., Giri, A. P. & Roy, R. A. Revised structure zone model for thin film physical structure. *Journal of Vacuum Science & Technology A: Vacuum, Surfaces, and Films* **2,** 500–503 (1984).

15. Barna, P. B. & Adamik, M. Fundamental structure forming phenomena of polycrystalline films and the structure zone models. *Thin Solid Films* **317,** 27–33 (1998).

16. Petrov, I., Barna, P. B., Hultman, L. & Greene, J. E. Microstructural evolution during film growth. *Journal of Vacuum Science & Technology A: Vacuum, Surfaces, and Films* **21,** S117-S128; 10.1116/1.1601610 (2003).

17. Mukherjee, S. & Gall, D. Structure zone model for extreme shadowing conditions. *Thin Solid Films* **527,** 158–163 (2013).

18. Mahieu, S., Ghekiere, P., Depla, D. & Gryse, R. de. Biaxial alignment in sputter deposited thin films. *Thin Solid Films* **515,** 1229–1249; 10.1016/j.tsf.2006.06.027 (2006).

19. Anders, A. A structure zone diagram including plasma-based deposition and ion etching. *Thin Solid Films* **518,** 4087–4090; 10.1016/j.tsf.2009.10.145 (2010).

20. Stein, H. *et al.* A structure zone diagram obtained by simultaneous deposition on a novel step heater. A case study for Cu 2 O thin films. *Phys. Status Solidi A* **212,** 2798–2804; 10.1002/pssa.201532384 (2015).

21. Bouaouina, B. *et al.* Nanocolumnar TiN thin film growth by oblique angle sputter-deposition. Experiments vs. simulations. *Materials & Design* **160,** 338–349; 10.1016/j.matdes.2018.09.023 (2018).

22. Wang, P., He, W., Mauer, G., Mücke, R. & Vaßen, R. Monte Carlo simulation of column growth in plasma spray physical vapor deposition process. *Surface and Coatings Technology* **335,** 188–197 (2018).

23. Savaloni, H. & Shahraki, M. G. A computer model for the growth of thin films in a structure zone model. *Nanotechnology* **15,** 311 (2003).

24. Müller, K. Stress and microstructure of sputter-deposited thin films. Molecular dynamics investigations. *Journal of Applied Physics* **62,** 1796–1799; 10.1063/1.339559 (1987).

25. Krüger, D. & Brinkmann, R. P. Interaction of magnetized electrons with a boundary sheath. Investigation of a specular reflection model. *Plasma Sources Sci. Technol.* **26,** 115009; 10.1088/1361-6595/aa9248 (2017).

26. Krüger, D., Trieschmann, J. & Brinkmann, R. P. Scattering of magnetized electrons at the boundary of low temperature plasmas. *Plasma Sources Sci. Technol.* **27,** 25011; 10.1088/1361-6595/aaaa85 (2018).

27. Trieschmann, J. *et al.* Combined experimental and theoretical description of direct current magnetron sputtering of Al by Ar and Ar/N 2 plasma. *Plasma Sources Sci. Technol.* **27,** 54003; 10.1088/1361-6595/aac23e (2018).

28. Music, D. *et al.* Correlative plasma-surface model for metastable Cr-Al-N. Frenkel pair formation and influence of the stress state on the elastic properties. *Journal of Applied Physics* **121,** 215108; 10.1063/1.4985172 (2017).

29. Music, D., Geyer, R. W. & Schneider, J. M. Recent progress and new directions in density functional theory based design of hard coatings. *Surface and Coatings Technology* **286,** 178–190; 10.1016/j.surfcoat.2015.12.021 (2016).

30. DeCost, B. L., Francis, T. & Holm, E. A. Exploring the microstructure manifold. Image texture representations applied to ultrahigh carbon steel microstructures. *Acta Materialia* **133,** 30–40; 10.1016/j.actamat.2017.05.014 (2017).

31. Kitahara, A. R. & Holm, E. A. Microstructure Cluster Analysis with Transfer Learning and Unsupervised Learning. *Integr Mater Manuf Innov* **7,** 148–156; 10.1007/s40192-018-0116-9 (2018).

32. Bulgarevich, D. S., Tsukamoto, S., Kasuya, T., Demura, M. & Watanabe, M. Pattern recognition with machine learning on optical microscopy images of typical metallurgical microstructures. *Scientific reports* **8,** 2078; 10.1038/s41598-018-20438-6 (2018).

33. Kondo, R., Yamakawa, S., Masuoka, Y., Tajima, S. & Asahi, R. Microstructure recognition using convolutional neural networks for prediction of ionic conductivity in ceramics. *Acta Materialia* **141,** 29–38; 10.1016/j.actamat.2017.09.004 (2017).

34. Chowdhury, A., Kautz, E., Yener, B. & Lewis, D. Image driven machine learning methods for microstructure recognition. *Computational Materials Science* **123,** 176–187; 10.1016/j.commatsci.2016.05.034 (2016).

35. Rovinelli, A., Sangid, M. D., Proudhon, H. & Ludwig, W. Using machine learning and a data-driven approach to identify the small fatigue crack driving force in polycrystalline materials. *npj Comput Mater* **4,** 963; 10.1038/s41524-018-0094-7 (2018).

36. Moot, T. *et al.* Material informatics driven design and experimental validation of lead titanate as an aqueous solar photocathode. *Materials Discovery* **6,** 9–16; 10.1016/j.md.2017.04.001 (2016).

37. Cao, B. *et al.* How To Optimize Materials and Devices via Design of Experiments and Machine Learning: Demonstration Using Organic Photovoltaics. *ACS nano* **12,** 7434–7444; 10.1021/acsnano.8b04726 (2018).

38. Salakhutdinov, R. Learning Deep Generative Models. *Annu. Rev. Stat. Appl.* **2,** 361–385; 10.1146/annurev-statistics-010814-020120 (2015).

39. Kingma, D. P. & Welling, M. Auto-encoding variational bayes. *arXiv preprint arXiv:1312.6114* (2013).

40. Goodfellow, I. *et al. in Advances in Neural Information Processing Systems 27,* edited by Z. Ghahramani, M. Welling, C. Cortes, N. D. Lawrence & K. Q. Weinberger (Curran Associates, Inc2014), pp. 2672–2680.

41. Stein, H. S., Guevarra, D., Newhouse, P. F., Soedarmadji, E. & Gregoire, J. M. Machine learning of optical properties of materials–predicting spectra from images and images from spectra. *Chemical science* **10,** 47–55 (2019).

42. Gomez-Bombarelli, R. *et al.* Automatic Chemical Design Using a Data-Driven Continuous Representation of Molecules. *ACS CENTRAL SCIENCE* **4,** 268–276; 10.1021/acscentsci.7b00572 (2018).

43. Yang, Z. *et al.* Microstructural Materials Design Via Deep Adversarial Learning Methodology. *Journal of Mechanical Design* **140**; 10.1115/1.4041371 (2018).

44. Li, X., Yang, Z., Brinson, L. C., Choudhary, A., Agrawal, A. & Chen, W. eds. *A Deep Adversarial Learning Methodology for Designing Microstructural Material Systems* (2018).

45. Noraas, R., Somanath, N., Giering, M. & Olusegun, O. O. in *AIAA Scitech 2019 Forum* (American Institute of Aeronautics and Astronautics2019).

46. Stueber, M., Diechle, D., Leiste, H. & Ulrich, S. Synthesis of Al–Cr–O–N thin films in corundum and f.c.c. structure by reactive r.f. magnetron sputtering. *Thin Solid Films* **519,** 4025–4031; 10.1016/j.tsf.2011.01.144 (2011).

47. Hofmann, S. & Jehn, H. A. Oxidation behavior of CrNx and (Cr, Al) Nx hard coatings. *Materials and Corrosion* **41,** 756–760 (1990).

48. Kunisch, C., Loos, R., Stüber, M. & Ulrich, S. Thermodynamic modeling of Al-Cr-N thin film systems grown by PVD. *Zeitschrift fur Metallkunde* **90,** 847–852 (1999).

49. Sugishima, A., Kajioka, H. & Makino, Y. Phase transition of pseudobinary Cr–Al–N films deposited by magnetron sputtering method. *Surface and Coatings Technology* **97,** 590–594 (1997).

50. Bobzin, K. *et al.* Mechanical properties and oxidation behaviour of (Al, Cr) N and (Al, Cr, Si) N coatings for cutting tools deposited by HPPMS. *Thin Solid Films* **517,** 1251–1256 (2008).

51. Schölkopf, B., Smola, A. & Müller, K.-R. eds. *Kernel principal component analysis* (Springer, 1997).

52. Castaldi, L. *et al.* Effect of the oxygen content on the structure, morphology and oxidation resistance of Cr–O–N coatings. *Surface and Coatings Technology* **203,** 545–549; 10.1016/j.surfcoat.2008.05.018 (2008).

53. Grochla, D. *et al.* Time- and space-resolved high-throughput characterization of stresses during sputtering and thermal processing of Al–Cr–N thin films. *J. Phys. D: Appl. Phys.* **46,** 84011; 10.1088/0022-3727/46/8/084011 (2013).

54. Mayrhofer, P. H., Music, D., Reeswinkel, T., Fuß, H.-G. & Schneider, J. M. Structure, elastic properties and phase stability of $Cr_{1-x}Al_xN$. *Acta Materialia* **56,** 2469–2475 (2008).

55. Bagcivan, N., Bobzin, K. & Theiß, S. $(Cr_{1-x}Al_x)$ N: a comparison of direct current, middle frequency pulsed and high power pulsed magnetron sputtering for injection molding components. *Thin Solid Films* **528,** 180–186 (2013).

56. Hultman, L., Sundgren, J., Greene, J. E., Bergstrom, D. B. & Petrov, I. High-flux low-energy ($\approx$ 20 eV) $N_2^+$ ion irradiation during TiN deposition by reactive magnetron sputtering: Effects on microstructure and preferred orientation. *Journal of Applied Physics* **78,** 5395–5403 (1995).

57. Hecimovic, A., Burcalova, K. & Ehiasarian, A. P. Origins of ion energy distribution function (IEDF) in high power impulse magnetron sputtering (HIPIMS) plasma discharge. *J. Phys. D: Appl. Phys.* **41,** 95203; 10.1088/0022-3727/41/9/095203 (2008).

58. X. Hou, L. Shen, K. Sun & G. Qiu eds. *Deep Feature Consistent Variational Autoencoder.* 2017 IEEE Winter Conference on Applications of Computer Vision (WACV) (2017).

59. Ledig, C. *et al. in Proceedings of the IEEE conference on computer vision and pattern recognition* (2017), pp. 4681–4690.

60. Mirza, M. & Osindero, S. Conditional generative adversarial nets. *arXiv preprint arXiv:1411.1784* (2014).

61. Tholander, C., Alling, B., Tasnádi, F., Greene, J. E. & Hultman, L. Effect of Al substitution on Ti, Al, and N adatom dynamics on TiN(001), (011), and (111) surfaces. *Surface Science* **630,** 28–40; 10.1016/j.susc.2014.06.010 (2014).

62. Bagcivan, N., Bobzin, K., Brögelmann, T. & Kalscheuer, C. Development of (Cr, Al) ON coatings using middle frequency magnetron sputtering and investigations on tribological behavior against polymers. *Surface and Coatings Technology* **260,** 347–361 (2014).

63. Reiter, A. E., Derflinger, V. H., Hanselmann, B., Bachmann, T. & Sartory, B. Investigation of the properties of Al1- xCrxN coatings prepared by cathodic arc evaporation. *Surface and Coatings Technology* **200,** 2114–2122 (2005).

64. Greczynski, G. *et al.* A review of metal-ion-flux-controlled growth of metastable TiAlN by HIPIMS/DCMS co-sputtering. *Surface and Coatings Technology* **257,** 15–25; 10.1016/j.surfcoat.2014.01.055 (2014).

65. Bohlmark, J. *et al.* The ion energy distributions and ion flux composition from a high power impulse magnetron sputtering discharge. *Thin Solid Films* **515,** 1522–1526; 10.1016/j.tsf.2006.04.051 (2006).

66. Horwat, D. & Anders, A. Spatial distribution of average charge state and deposition rate in high power impulse magnetron sputtering of copper. *J. Phys. D: Appl. Phys.* **41,** 135210; 10.1088/0022-3727/41/13/135210 (2008).

67. Banko, L. *et al.* Effects of the ion to growth flux ratio on the constitution and mechanical properties of Cr1-x-Alx-N thin films. under review ACS Combinatorial Science (2019).

68. Zuiderveld, K. in *Graphics gems IV* (1994), pp. 474–485.